\documentclass{article}%
\usepackage{amsmath}%
\usepackage{amsfonts}%
\usepackage{amssymb}%
\usepackage{graphicx}

\providecommand{\U}[1]{\protect\rule{.1in}{.1in}}

\begin{document}

\title{Evolution of quasi-history in a Physics Textbook}
\author{J.R. Persson\\Norwegian University of Science and Technology\\N-7491 Trondheim \\Norway}
\maketitle

\section{Introduction}

The primary aim, in teaching physics, is that the student should gain
an understanding of the principles of physics and how to apply them to
different problems. A secondary aim is to allow the students to appreciate the
scientific approach and significance of it in the evolution of science and
society. One approach for the second aim has been to include "historical
material" in physics textbooks. The quantity of the historical material
included is quite diverse, from textbooks with a very strong historical
approach \ to others without any historical material. The quality of the
material included is also diverse. In this article we focus on the development
of the historical material, i.e. a certain historical development, in a
specific textbook (Sears \& Zemansky's University Physics) over a number of
editions. The aim is to see when and how the historical material is included
and how well it describes the actual history. Will the physics adapt to
history or vice-verse.

The event of interest is the introduction of the Blackbody radiation formula
or Planck's radiation formula. This is well known out of an historical
perspective, but also a case where a quasi-historical is quite common in textbooks.

\section{ Planck's radiation formula}

There exist excellent accounts of the development of quantum theory
in general \cite{Mehra82}\cite{Jammer66}\cite{Herman71} as well as detailed
studies of Planck's radiation formula \cite{Klein62}\cite{Klein63}%
\cite{Kragh00}, why we will not go into detail on the history, but try to give
a general outline.

The development of spectroscopy made it possible to study spectra from
different elements but also from macroscopic objects. In 1859 argued Robert
Kirchoff that black-body radiation was of a fundamental nature, thus
initiating an interest in these studies. The spectral distribution were
investigated by several physicists, both experimentalists and theorists. In
1896 Wilhelm Wien found a radiation law that was in agreement with precise
measurements performed at the Physikalisch-Technische Reichsanstalt in Berlin.

According to Wien the spectral density is described as:%

\[
u\left(  f,T\right)  =\alpha f^{3}e^{\frac{-\beta f}{T}}%
\]

where $\alpha$ and $\beta$ are constants to be determined empirically, $f$ and
$T$ are frequency and temperature, respectively. However, this lacked a
theoretical foundation, something that Max Planck could not accept. Planck
formulated a "principle of elementary disorder", which he used to define the
entropy of an ideal oscillator. Using this it was possible for Planck, early in
1899, to find an expression from which Wien's law followed.

However, measurements by Lummer and Pringsheim in November 1899, showed a
deviation from Wien's law at low frequencies. Planck had to revise his
calculations, using a new expression for the entropy of a single oscillator.
The new distribution law was presented at a meeting of the German Physical
Society on 19 October 1900. The spectral density was now given as:%

\[
u\left(  f,T\right)  =\frac{\alpha f^{3}}{e^{\frac{\beta f}{T}}-1}%
\]

where $\alpha$ and $\beta$ are constants to be determined
empirically, $f$ and $T$ are frequency and temperature, respectively. The
problem was, as Planck realized, that his new expression was no more than an
inspired guess. Planck had not used energy quantization nor Boltzmann's
probabilistic interpretation of entropy.

In what he himself describes as "an act of desperation", he turned to
Boltzmann's probabilistic notion of entropy. Even if he adopted Boltzmann's
view, he did not convert to the probabilistic notion of entropy. He remained
convinced that the law of entropy was absolute, not probabilistic, and
therefore reinterpreted Boltzmann's theory in his own non-probabilistic way.
Using the "Boltzmann equation"

\[
S=k\log W
\]

, which relates the entropy, $S$, to the molecular disorder, $W$. In order to
determine $W$, Planck had to be able to count the number of ways a given amount of energy
can be distributed in a set of oscillators. It was in doing this Planck
introduced what he called "energy elements", that is the total energy of
the black-body oscillators, E, divided into finite portions of energy,
$\varepsilon$, via a process known as "quantization". The energy of the finite
energy elements were given by a constant, $h$, multiplied with a frequency $f$.

\[
\varepsilon=hf
\]

Using this it was easy for Planck to follow the procedure that Boltzmann used
in deriving Maxwell's distribution of velocities of molecules in a gas and
derive the spectral density:\bigskip

\[
u\left(  f,T\right)  =\frac{8\pi}{c^{3}}\frac{hf^{3}}{\left(  e^{\frac{hf}%
{kT}}-1\right)  }%
\]

This was presented to the German Physical Society on 14 December
1900 \cite{Planck1900}, followed by four papers in 1901.

Lord Rayleigh published a paper in June 1900 \cite{Rayleigh00} where he
presented an improved version of Wien's radiation law. Using
Maxwell-Boltzmann's equipartition theorem, he obtained a different radiation law:

\[
u\left(  \lambda,T\right)  =c_{1}\frac{T}{\lambda^{4}}e^{\frac{-c_{2}}{\lambda
T}}%
\]

This law was noted and tested by experimentalists, but got very little
attention since Planck had produced a formula that was a better fit to the
experimental results. In 1905 came Rayleigh \cite{Rayleigh1905} with a refined
radiation law:%

\[
u\left(  \lambda,T\right)  =\frac{64\pi kT}{\lambda^{4}}%
\]

An error in the calculations was corrected by Jeans\cite{Jeans1905}%
\cite{Jeans1905a} and the new radiation law was therefore named Rayleigh-Jeans law:%

\[
u\left(  \lambda,T\right)  =\frac{8\pi kT}{\lambda^{4}}%
\]

The result is an energy density that increases as the frequency gets higher,
becoming "catastrophic" in the ultraviolet region. This made Paul Ehrenfest
coin the name "ultraviolet catastrophe" in 1911, thus becoming a matter of
discussion quite late in the development of quantum theory.

Einstein was the first to fully adopt the quantisation principle in his
derivation of Planck's radiation law in 1906 \cite{Einstein06}. Something that caused
an increased interest among physicists, leading to more discussions.

The time scale is quite clear: Planck's Radiation law origins from 1900,
Rayleigh-Jeans law from 1905 and the term "ultraviolet catastrophe" from 1911.

From a physical point of view, the order of the radiation laws will be
different. Based on classical physics, i.e. no quantization, Rayleigh-Jeans law
should be placed before Planck's law. Also the use of the Maxwell-Boltzmann's
equipartition theorem, makes it natural to derive it after the
Maxwell--Boltzmann distribution.

\section{ Sears \& Zemansky's University Physics}

The textbook chosen for this study, is Sears \& Zemansky's University Physics
with the first edition  published in 1949 \cite{Sears1949} and with later
editions widely used around the world. At many universities this textbook is
the first (and sometimes the only)\ choice, leaving a rather large impact on
students. Since this textbook has a long history will it be the obvious choice
when studying the developments of textbooks. Since the first edition in
1949 \cite{Sears1949}, a number of editions has been published. The responsible
authors have changed over the years. The first four editions,
1949 \cite{Sears1949}, 1955 \cite{Sears1955}, 1963 \cite{Sears1963} and
1970 \cite{Sears1970} were written by Sears and Zemansky. The fifth edition,
from 1976 \cite{Sears1976}, was coauthored by Young. This edition was also the
last coauthored with Sears who died suddenly during the final preparing stages
of the manuscript. Six years later, in 1982 \cite{Sears1982}, the sixth edition
was published, this being the last with Zemansky who died in 1981 after the
new manuscript had been finished. The seventh and eight editions were authored
by Young, 1987 \cite{Young1987} and 1992 \cite{Young1992}, respectively. The
ninth edition included Freedman as co-author, something that has been the case
for the last editions up to date, ninth edition in 1996 \cite{Young1996}, tenth to
thirteenth, 2000 \cite{Young2000}, 2004 \cite{Young2004}, 2008 \cite{Young2008},
and 2012 \cite{Young2012}, respectively. \newline It should also be mentioned
that a number of contributing authors has been used over the years, further
informations on these can be found in the preface of the different
editions.\newline Having a textbook that has evolved from the start over 60
years ago, gives an unique opportunity to probe the development of the teaching of
physics. Both in the presentation and the layout, but also in the examples and
derivations used. In this case we are more interested in the development of
how a specific historical development is presented and how it is changing. One
have to remember that University Physics was not written in a vacuum, there
existed other textbooks before, and Sears had written a book before. The first
edition followed the a basic outline, with a conventional selection and
sequence. Mechanics, heat, sound, electricity and magnetism, and optics.
Modern Physics were added after the first editions, with atomic physics as the
first modern subject to be included. It is clearly stated in the preface of
the first editions (up to the seventh) that:

\begin{quotation}
The emphasis is on physical principles; historical background and practical
applications have been given a place of secondary importance.\cite[Preface]%
{Sears1963}
\end{quotation}

The eighth edition \cite{Young1992} is a comprehensive revision aimed at the
changes in the background and needs of the students as well as a change in the
philosophy of introductionary physics courses. One effect of this being an
attempt to make physics more human, in part by including the history of physics.
One should note that the third edition \cite{Sears1963} is special, as it marks an increased difficulty in the presentation
as well as introduction of new approaches. However, the fourth edition \cite{Sears1970} is a step back to the earlier presentations and a slight reduction in the mathematical difficulty.

\subsection{Blackbody radiation in University Physics.}

In the first and second editions the Blackbody radiation is described from an experimental radiation view, without mentioning Planck's quantisation.
 Blackbody radiation and Planck's quantisation is mentioned in the
third edition of University Physics, in the thermodynamics section (Chapter
17-7 Planck's law \cite{Sears1963}). The description differs from earlier
editions, with a discussion on the origin of Planck's law, as the earlier
editions just described Blackbody radiation from a radiation point of view
without mentioning Planck's quantisation. The presentation in the third edition, does
not emphasize on the history, but on a more experimentally observed approach:

\begin{quotation}
Max Planck, in 1900, developed an empirical equation that satisfactory
represented the observed energy distribution in the spectrum of a Blackbody.
After unsuccessful attempts to justify his equation by theoretical reasoning
based on the laws of classical physics, Planck concluded that these laws did
not apply to energy transformations on an atomic scale. Instead, he postulated
that a radiating body consisted of an enormous number of elementary
oscillators, some vibrating at one frequency and some at another, with all
frequencies from zero to infinity being represented.\cite[p 380]{Sears1963}
\end{quotation}

The discussion in the fourth, fifth and sixth editions, is very similar to the discussion in the first two edidtions, making the third edition special.

It is with the seventh edition (1987) that Planck makes his entrance in a
chapter on continuous spectra (Chapter 41-7 \cite{Young1987}). The
presentation takes a starting point in the empirical Stefan-Boltzmann Law and
the Wien displacement law, and describes how Planck in 1900 used the principle
of equipartition of energy together with a quantisation of the energy that is
emitted or absorbed. The presentation does not emphasise the history but states:

\begin{quotation}
Finally, in 1900 Max Planck succeeded in deriving a function, now called the
Planck radiation law, that agreed with experimentally obtained
power-distribution curves. To do this he added to the classical equipartition
theorem the additional assumption that a harmonic oscillator with frequency f
can gain or lose energy only in discrete steps of magnitude hf, where h is the
same constant that now bears his name. \newline Ironically, Planck himself
originally regarded this quantum hypothesis, as a calculational trick rather
than a fundamental principle. But as we have seen, evidence for the quantum
aspects of light accumulated, and by 1920 there was no longer any doubt about
the validity of the concept. Indeed, the concept of discrete energy levels of
microscopic systems really originated with Planck, not Bohr, and we have
departed from the historical order of things by discussing atomic spectra
before continuous spectra.\cite[p995]{Young1987}
\end{quotation}

It is notable that there is no discussion on the development of physics, it is
just the result that is important. The description differs from the historical
development but can not be considered as quasi-history.

\subsection{Changes in the eight edition.}

The eight edition (1992)\cite{Young1992} show some major changes, the title
now includes modern physics, and the material in this section has changed a
lot and new material has been added. In the case of continuous spectra
(Chapter 40-8) , the discussion about the physics behind has been modernised,
but the history has also been extended with the derivation of Rayleigh-Jeans
law from 1905, presented as a precursor to Planck's Law from 1900!

\begin{quotation}
During the last decade of the nineteenth century, many attempts were made to
\textit{derive} these empirical results from basic principles. In one attempt,
Rayleigh considered light enclosed in a rectangular box with perfectly
reflecting sides. Such a box has a series of possible normal modes for the
light waves, analogous to \textit{normal modes} for a string held at both ends
(Section 20- 3). It seemed reasonable to assume that the distribution of
energy among the various modes was given by the equipartition principle
(Section 16- 4), which had been successfully used in the analysis of heat
capacities. A small hole in the box would behave as an ideal Blackbody
radiator.\newline Including both the electric- and magnetic-field energies,
Rayleigh assumed that the total energy of each normal mode was equal to kT.
Then by computing the number of normal modes corresponding to a wavelength
interval $d\lambda$ , Rayleigh could predict the distribution of wavelengths
in the radiation within the box. Finally, he could compute the intensity
distribution $I\left(  \lambda\right)  $ of the radiation emerging from a
small hole in the box. His result was quite simple:

\hspace{5mm} \hspace{5mm} $I\left(  \lambda\right)  =\frac{2\pi ckT}%
{\lambda^{4}}\qquad$(40-30)

At large wavelengths this formula agrees quite well with the experimental
results shown in Fig. 40- 27, but there is serious disagreement at small
$\lambda$. The experimental curve falls to zero at small $\lambda$; Rayleigh's
curve approaches infinity as $1/\lambda^{4}$, a result called in Rayleigh's
time the {"}ultraviolet catastrophe.{"} Even worse, the integral of Eq. (40-
30) over all $\lambda$ is infinite, indicating an infinitely large
\textit{total} radiated intensity. Clearly, something is wrong.\newline
Finally, in 1900, Planck succeeded in deriving a function, now called the
\textbf{Planck radiation law}, that agreed very well with experimental
intensity distribution curves. To do this, he made what seemed at the time to
be a crazy assumption; he assumed that in Rayleigh's box a normal mode with
frequency $f$ could gain or lose energy only indiscrete steps with magnitude $hf$,
where $h$ is the same constant that now bears Planck's name.\cite[p1129-1130]%
{Young1992}
\end{quotation}

Note the sharp wording in connection to Planck's assumption.\newline We also
find a change in the description of development of physics, which is now a
part of the discussion. This indicates a desire to discuss how the results
were obtained. \ But we can also notice that the description of events does
not follow the actual history, in which Rayleigh plays a minor part.

\begin{quotation}
Planck was not comfortable with this \textit{quantum hypothesis}; he regarded
it as a calculational trick rather than a fundamental principle. In a letter
to a friend he called it {"}an act of desperation{"} into which he was forced
because {"}a theoretical explanation had to be found at any cost, whatever the
price.{"} But five years later, Einstein extended this concept to explain the
photoelectric effect (Section 40- 2), and other evidence quickly mounted. By
1915 there was no longer any doubt about the validity of the quantum concept
and the existence of photons. By discussing atomic spectra \textit{before}
discussing continuous spectra, we have departed from the historical order of
things. The credit for inventing the concept of quantization goes to Planck,
even though he didn't believe it at first.\cite[p1130]{Young1992}
\end{quotation}

Note that the year for acceptance of quantisation has changed from 1920 to
1915! \newline It is clear that the style of the book has changed in such a
way that it is not sufficient to just present the results but the development
on how the results was found is becoming more important.

The ninth (1996)\cite{Young1996}, tenth (2000)\cite{Young2000} and eleventh
(2004)\cite{Young2004} editions are almost the same as the eight, with minor
changes in the presentation.

\subsection{ Changes in the twelfth edition.}

In the twelfth edition (2008)\cite{Young2008} the change is not in the
presentation but in the fact that Rayleigh's derivation is now given as a
separate subsection with a special heading, \textbf{Rayleigh and the
"Ultraviolet Catastrophe"}{, in section 38.8. }As before placed before
Planck's derivation which also is given a separate subsection with a special
heading, \textbf{Planck and the Quantum Hypothesis}. With the presentation and
now subsection the quasi-history is enhanced, leading further away form the
actual events.

\subsection{The thirteenth edition (2012)}

The presentation in the thirteenth edition \cite{Young2012} is the same when
it comes to history. But now an analogy is introduced to explain the
difference between line spectra and continuous spectra. In Planck's
derivation, the discussion is extended and coupled more towards thermodynamics
and Boltzmann distributions.

\section{Discussion}

It is clear that the presentation style changed from a result-centered
approach to a broader approach as from the eighth edition, where the idea to
make it more interesting but introducing a human dimension by telling the
history. \ The aim is to present a logical presentation of the scientific
facts, but also to provide a historical framework inside which the scientific
facts fit easily. Thus creating a development of physics that make sense and
is easily remembered. In this process, the historical events have to adapt to
the physics, in such way that students could follow a "logical" development.
However, the history is seldom "logical", so that it is  easy to rewrite
history to fit the physics. In doing so one misses the chance to discuss the
foundations of physics and why these might seem to be counterintuitive. It is
important to show that understanding and finding the correct theory is seldom
straightforward. If the students have trouble to understand a theory it might
be comforting to know that the persons developing the theory also had problems
in understanding it.

The textbook studied, does not follow the historical development, but tend to
enhance a quasi-historical myth, when placing a reasonable results-centered
presentation in a false historical context. The intentions and the pedagogical
motivation can be considered to be solid, but unfortunately this gives a false
picture of the history and nature of physics.


\begin{thebibliography}{10}                                                                                               %
\bibitem {Mehra82}Mehra, J., \& Rechenherg, H. (1982). The Historical
Development of Quantum Theory (Vol.1). New York: Springer-Verlag.

\bibitem {Jammer66}Jammer, M. (1966). Conceptual Development of Quantum
Mechanics . New York: McGraw-Hill.

\bibitem {Herman71}Herman, A., (1971) The Genesis of Quantum Theory
(1899-1913). Cambridge. Mass.:MIT Press.

\bibitem {Klein62}Klein, M. J. (1962). Max Planck and the Beginning of the
Quantum Theory. Archive for History of Exact Sciences, 1(5), 459-479.

\bibitem {Klein63}Klein, M. J. (1963). Einstein's First Paper on Quanta. The
Natural Philosopher, 2, 59-86.

\bibitem {Kragh00}Kragh, H. (2000). Max Planck: the reluctant revolutionary.
Physics World December 2000.

\bibitem {Planck1900}Planck,M., (1900). Zur Theorie des Gesetzes der
Energieverteilung im Normalspectrum. Verhandl. Dtsch. phys. Ges., 2, 237-245

\bibitem {Rayleigh00}Rayleigh, L. (1900). Remarks upon the Laws of Complete
Radiation. Philosophical Magazine, 49(5), 539-540.

\bibitem {Rayleigh1905}Rayleigh, L. (1905). The Dynamical Theory of Gases and
of Radiation. Nature, 72, 54.

\bibitem {Jeans1905}Jeans,J., (1905). The partition of energy between matter
and ether. Phil. Mag. 10,91-98.

\bibitem {Jeans1905a}Jeans,J., (1905). The law of radiation. Proc.Roy.Soc. A, 76,545-552

\bibitem {Einstein06}Einstein, A. (1906). Zur Theorie der Lichterzeugung und
Lichtabsorption. Annalen der Physik, 20(4), 199-207.

\bibitem {Sears1949}Sears, F.W., \& Zemansky, M.W., (1949). University Physics
(1 st.). Cambridge Mass.: Addison Wesley.

\bibitem {Sears1955}Sears, F.W., \& Zemansky, M.W., (1955). University Physics
(2 nd.). Reading Mass.: Addison Wesley.

\bibitem {Sears1963}Sears, F.W., \& Zemansky, M.W., (1963). University Physics
(3 ed.). Reading Mass.: Addison Wesley.

\bibitem {Sears1970}Sears, F.W., \& Zemansky, M.W., (1970). University Physics
(4 ed.). Reading Mass.: Addison Wesley.

\bibitem {Sears1976}Sears, F.W., Zemansky, M.W., \& Young, H. D. (1976).
University Physics (5 ed.). Reading Mass.: Addison Wesley.

\bibitem {Sears1982}Sears, F.W., Zemansky, M.W., \& Young, H. D. (1982).
University Physics (6 ed.). Reading Mass.: Addison Wesley.

\bibitem {Young1987}Young, H. D. (1987). University Physics (7 ed.). Reading
Mass.: Addison Wesley.

\bibitem {Young1992}Young, H. D. (1992). University Physics, Extended version
with modern physics (8 ed.). Reading Mass.: Addison Wesley.

\bibitem {Young1996}Young, H. D., \& Freedman, R. A. (1996). University
Physics (9 ed.). San Fransisco Boston New York: Addison Wesley.

\bibitem {Young2000}Young, H. D., \& Freedman, R. A. (2000). University
Physics (10 ed.). San Fransisco Boston New York: Addison Wesley.

\bibitem {Young2004}Young, H. D., \& Freedman, R. A. (2004). University
Physics (11 ed.). San Fransisco Boston New York: Pearson Addison Wesley.

\bibitem {Young2008}Young, H. D., \& Freedman, R. A. (2008). University
Physics (12 ed.). San Fransisco Boston New York: Pearson Addison Wesley.

\bibitem {Young2012}Young, H. D., \& Freedman, R. A. (2012). University
Physics (13 ed.). San Fransisco Boston New York: Pearson Addison Wesley.
\end{thebibliography}
\end{document}